\documentclass[a4paper,10pt]{article}

\usepackage[dvips]{graphicx}
\usepackage{epsfig,amsmath,amssymb,verbatim,mathrsfs,array,layout,textcomp,amssymb,latexsym}

\newcommand{\beq}{\begin{eqnarray}}
\newcommand{\eeq}{\end{eqnarray}}

\def\beqa{\begin{eqnarray}}
\def\eeqa{\end{eqnarray}}
\newcommand{\no}{\nonumber}
\newcommand{\bv}{\left(\begin{array}{c}}
\newcommand{\ev}{\end{array}\right)}
\newcommand{\bmtwo}{\left(\begin{array}{cc}}
\newcommand{\bmthree}{\left(\begin{array}{ccc}}
\newcommand{\emn}{\end{array}\right)}
\newcommand{\bmtwoc}{\left\{\begin{array}{cc}}
\newcommand{\bmthreec}{\left\{\begin{array}{ccc}}
\newcommand{\emnc}{\end{array}\right\}}
\newcommand{\ba}{\begin{array}}
\newcommand{\ea}{\end{array}}

\def\lsim{\mathrel{\rlap{\lower4pt\hbox{\hskip1pt$\sim$}}
     \raise1pt\hbox{$<$}}}         
\def\gsim{\mathrel{\rlap{\lower4pt\hbox{\hskip1pt$\sim$}}
     \raise1pt\hbox{$>$}}}         

\begin{document}

\begin{titlepage}

\vskip1.5cm
\begin{center}
  {\Large \bf What if ${\rm BR}(h\to\mu\mu)/{\rm BR}(h\to\tau\tau)\neq m_\mu^2/m_\tau^2$?}
\end{center}
\vskip0.2cm

\begin{center}
Avital Dery, Aielet Efrati, Yonit Hochberg and Yosef Nir\\
\end{center}
\vskip 8pt

\begin{center}
{\it Department of Particle Physics and Astrophysics\\
Weizmann Institute of Science, Rehovot 76100, Israel} \vspace*{0.3cm}

{\tt  avital.dery,aielet.efrati,yonit.hochberg,yosef.nir@weizmann.ac.il}
\end{center}

\vglue 0.3truecm

\begin{abstract}
  \vskip 3pt \noindent Measurements of the Yukawa couplings of the recently discovered boson $h$ to fermion pairs will provide a new arena for studying flavor physics.
  We analyze the lessons that can be learned by measuring the $h$ decay rates into the charged lepton pairs, $\tau^+\tau^-$, $\mu^+\mu^-$ and $\tau^\pm\mu^\mp$. We demonstrate how this set of measurements can distinguish in principle between various classes of flavor models such as natural flavor conservation, minimal flavor violation, and Froggatt-Nielsen symmetry.
\end{abstract}

\end{titlepage}

\section{Introduction}
A Higgs-like boson $h$  has been discovered by the ATLAS and CMS experiments at the LHC~\cite{:2012gk,:2012gu}. The fact that for the $f=\gamma\gamma$ and $f=ZZ^*$ final states, the experiments measure
\beq\label{eq:defmu}
R_f\equiv\frac{\sigma(pp\to h){\rm BR}(h\to f)}
{[\sigma(pp\to h){\rm BR}(h\to f)]^{\rm SM}},
\eeq
of order one (see {\it e.g.}~\cite{Carmi:2012zd}),
\beqa\label{eq:muexp}
R_{ZZ^*}&=&1.0\pm0.4,\\
R_{\gamma\gamma}&=&1.6\pm0.3,
\eeqa
is suggestive that the $h$-production via gluon-gluon fusion proceeds at a rate similar to the Standard Model (SM) prediction. ATLAS finds the ratio of the gluon-gluon coupling to the higgs-like particle normalized to the SM value to be $1.1^{+1.4}_{-0.2}$~\cite{ATLAScoupling}. This gives a strong indication that $Y_t$, the $ht\bar t$ Yukawa coupling, is of order one. This first determination of $Y_t$ signifies a new arena for the exploration of {\it flavor physics}.

In the future, measurements of $R_{b\bar b}$ and $R_{\tau^+\tau^-}$ will allow us to extract additional flavor parameters: $Y_b$, the $hb\bar b$ Yukawa coupling, and $Y_\tau$, the $h\tau^+\tau^-$ Yukawa coupling. For the latter, the current allowed range, obtained from combining the ATLAS~\cite{ATLAStau} and CMS~\cite{CMStau} results, is already quite restrictive:
\beq
R_{\tau^+\tau^-}=0.7\pm0.4.
\eeq
It may well be that the values of $Y_b$ and/or $Y_\tau$ will deviate from their SM values. The most likely explanation of such deviations will be that there are more than one Higgs doublets, and that the doublet(s) that couple to the down and charged lepton sectors are not the same as the one that couples to the up sector.

A more significant test of our understanding of flavor physics, which might provide a window into new flavor physics, will come further in the future,
when $R_{\mu^+\mu^-}$ is measured. The ratio
\beq\label{eq:rmutau}
X_{\mu^+\mu^-}\equiv\frac{{\rm BR}(h\to\mu^+\mu^-)}{{\rm BR}(h\to\tau^+\tau^-)},
\eeq
is predicted within the SM with impressive theoretical cleanliness. To leading order, it is given by $X_{\mu^+\mu^-}=m_\mu^2/m_\tau^2$, and the corrections of order $\alpha_W$ and of order $m_\mu^2/m_\tau^2$ to this leading result are known (see Eq. (\ref{eq:xmmsm}) below). The main question that we analyze in this work is what can be learned from a test of this relation.

It is also possible to search for the SM-forbidden decay modes, $h\to\mu^\pm\tau^\mp$, see {\it e.g.} \cite{Han:2000jz,Cotti:2001fm,Assamagan:2002kf,Arcelli:2004af,Goudelis:2011un,Blankenburg:2012ex,Harnik:2012pb,Davidson:2012ds,Arhrib:2012ax,Chiang:2013}. A measurement of, or an upper bound on
\beq\label{eq:brhmt}
X_{\mu\tau}\equiv\frac{{\rm BR}(h\to\mu^+\tau^-)+{\rm BR}(h\to\mu^-\tau^+)}{{\rm BR}(h\to\tau^+\tau^-)},
\eeq
would provide additional information relevant to flavor physics. Thus, a broader goal of this work is to understand the implications for flavor physics of measurements of $R_{\tau^+\tau^-}$, $X_{\mu^+\mu^-}$ and $X_{\mu\tau}$.

The plan of this paper is as follows. In Section \ref{sec:sm} we review the SM prediction for $X_{\mu^+\mu^-}$. The experimental constraints on the leptonic Yukawa couplings are presented in Section \ref{sec:exp}, first from indirect (loop) measurements (subsection \ref{sec:ind}), and then from collider searches (subsection \ref{sec:dir}). To demonstrate the power of measuring $R_{\tau^+\tau^-}$, $X_{\mu^+\mu^-}$, and $X_{\tau\mu}$ in probing flavor models, we obtain the predictions of four different models of new physics for these observables: multi-Higgs doublet models with natural flavor conservation (Section \ref{sec:nfc}), a single-Higgs model with non-renormalizable terms that are subject to the principle of minimal flavor violation (Section \ref{sec:mfv}) or to selection rules from a Froggatt-Nielsen symmetry (Section \ref{sec:fn}), and a model where the hierarchy in the Yukawa couplings comes from their Higgs dependence (Section \ref{sec:gl}). We conclude in Section \ref{sec:con}.

\section{SM predictions}
\label{sec:sm}
The calculation of the Higgs {\it decay rates} into dileptons is
rather accurate. Ref.~\cite{Denner:2011mq} estimates at the $1-2\%$ level the uncertainty on the decay rates from missing higher orders.
What is measured, however, are {\it branching ratios} (times
production rates). These are affected by the uncertainties on the
leading decay modes, particularly $h\to b\bar b$ and $h\to c\bar c$,
which involve the parametric uncertainties in $\alpha_s$, $m_c$, $m_b$
and $m_t$. For $m_h=125.5$ GeV, Ref. \cite{Denner:2011mq} quotes
\beq\label{eq:mmtt}
{\rm BR}(h\to\tau^+\tau^-)=6.24\times10^{-2}\times(1.000\pm0.057),\nonumber\\
{\rm BR}(h\to\mu^+\mu^-)=2.17\times10^{-4}\times(1.000\pm0.059),
\eeq
where the leading uncertainties are parametric. When we take the ratio
of the two branching ratios, however, it equals the ratio between the
decay rates, and these parametric uncertainties are irrelevant. (In
addition, the production rates cancel out.) What is left are the
$m_\ell$-dependent correction terms to the rates.

As concerns the uncertainties in the lepton masses, we have~\cite{PDG}
\beq\label{eq:mell}
m_\tau&=&1776.82\pm0.16\ {\rm MeV},\nonumber\\
m_\mu&=&105.6583715\pm0.0000035\ {\rm MeV}.
\eeq
Thus, the parametric uncertainty in the leading order relation,
$X_{\mu^+\mu^-}=(m_\mu/m_\tau)^2$, comes from the
uncertainty on $m_\tau$ and is of order $2\times10^{-4}$.

Within the SM, in lowest order,
\beq
\Gamma^{\rm LO}(h\to\ell^+\ell^-)=\frac{G_F m_h}{4\sqrt2\pi}m_\ell^2\beta_\ell^3,
\eeq
where
\beq
\beta_\ell=(1-4m_\ell^2/m_h^2)^{1/2}.
\eeq
The electroweak correction is approximated by~\cite{Bardin:1990zj,Dabelstein:1991ky, Kniehl:1991ze,Kniehl:1993ay,Djouadi:2005gi}
\beq
\Gamma(h\to\ell^+\ell^-)=\Gamma^{\rm LO}\left(1+\delta^\ell_{\rm QED}\right)\left(1+\delta^\ell_{\rm weak}\right),
\eeq
with
\beq\label{eq:del}
\delta^\ell_{\rm QED}&=&\frac{3\alpha}{2\pi}\left(\frac32-\log\frac{m_h^2}{m_\ell^2}\right)\,,\no\\
\delta^\ell_{\rm weak}&=&\frac{G_F}{8\pi^2\sqrt2}\left[7m_t^2+m_W^2\left(-5+\frac{3\log
      c_W^2}{s_W^2}\right)
  -m_Z^2\frac{6(1-8s_W^2+16s_W^4)-1}{2}\right].
\eeq
In the ratio $X_{\mu^+\mu^-}$ the weak corrections cancel out. The non-universal corrections that we do not include are of order $(\alpha/\pi)(m_\tau^2-m_\mu^2)/m_h^2\sim 10^{-6}$, smaller than the phase space factor by $\alpha$, and smaller than those included in $\delta^\ell_{\rm QED}$ by $m_\ell^2/m_h^2$.

Thus, for the ratio, we obtain
\beq
X_{\mu^+\mu^-}=\frac{m_\mu^2}{m_\tau^2}
\frac{\beta_\mu^3}{\beta_\tau^3}
\frac{1+\delta_{\rm QED}^\mu}{1+\delta_{\rm QED}^\tau}.
\eeq
Expanding to leading order in $m_\ell^2/m_h^2$, we find
\beq\label{eq:xmmsm}
X_{\mu^+\mu^-}=\frac{m_\mu^2}{m_\tau^2}
\left[1+\frac{6(m_\tau^2-m_\mu^2)}{m_h^2}
  -\frac{3\alpha}{2\pi}\log\frac{m_\tau^2}{m_\mu^2}\right].
\eeq
Given $m_h=125.5\ {\rm GeV}$, and $m_{\mu,\tau}$ of Eq. (\ref{eq:mell}), the value of the phase space factor is
\beq\label{eq:phsp}
\frac{6(m_\tau^2-m_\mu^2)}{m_h^2}=+0.0012.
\eeq
Using, in addition, $\alpha=1/137.0359997$, the value of the electroweak correction factor is
\beq\label{eq:ewmass}
-\frac{3\alpha}{2\pi}\log\frac{m_\tau^2}{m_\mu^2}=-0.020.
\eeq
Thus, the phase space factor is of order one permil, while the electroweak correction factor is of order two percent.

The running lepton masses $m_\ell(\mu)$ are related to the pole mass $m_\ell$ via \cite{Arason:1991ic}
\beq
m_\ell(\mu)=m_\ell\left\{1-\frac{\alpha(\mu)}{\pi}
  \left[1+\frac32\ln\frac{\mu}{m_\ell(\mu)}\right]\right\}.
\eeq
Thus,
\beq
\frac{m_\mu(m_h)}{m_\tau(m_h)}=\frac{m_\mu}{m_\tau}
\left(1-\frac{3\alpha(\mu)}{2\pi}\ln\frac{m_\tau}{m_\mu}\right).
\eeq
We learn that the electroweak correction (\ref{eq:ewmass}) can be
absorbed by changing from the pole masses to the running masses,
{\it i.e.}
\beq\label{eq:smxmm}
X_{\mu^+\mu^-}=\left[\frac{m_\mu(m_h)}{m_\tau(m_h)}\right]^2
\left[1+\frac{6(m_\tau^2-m_\mu^2)}{m_h^2}\right].
\eeq

In the above discussion of corrections to the rates, Eq.~\eqref{eq:del}, and thus to $X_{\mu^+\mu^-}$, Eq.~\eqref{eq:smxmm}, we have implicitly assumed that the measured rates to $\mu^+\mu^-$ and $\tau^+\tau^-$ are fully inclusive of bremmstrahlung radiation. In practice, experimental cuts on real photon emission are likely to be applied, introducing an important correction factor for each mode compared to the fully inclusive rates. This correction factor can be large, potentially of order $\sim 10\%$ (see {\it e.g.}~\cite{Buras:2012ru} for the effect of soft photon radiation on the $B\to \mu^+\mu^-$ decay), and can be different for the muon and tau modes, and might be above the other subleading corrections presented above. The precise evaluation of this effect is beyond the scope of this paper, and will be explored in future work.

Finally, we quote the value of the relevant Yukawa couplings in the SM:
\beqa\label{eq:smyukawa}
Y_{\tau}^{\rm SM}&=&1\times10^{-2},\nonumber\\
Y_{\mu}^{\rm SM}&=&6\times10^{-4},\nonumber\\
Y_{\mu\tau}^{\rm SM}=Y_{\tau\mu}^{\rm SM}&=&0.
\eeqa

The SM predictions for the Higgs decays into lepton pairs are presented in the SM row in Table \ref{tab:rxx}. For simplicity of presentation we write in the Table $X_{\mu^+\mu^-}^{\rm SM}=m_\mu^2/m_\tau^2$ instead of the more exact expression of Eq. (\ref{eq:smxmm}).

\begin{table}[t]
\caption{Predictions for $R_{\tau^+\tau^-}, X_{\mu^+\mu^-},\  X_{\mu^\pm\tau^\mp}$}
\label{tab:rxx}
\begin{center}
\begin{tabular}{cccc} \hline\hline
\rule{0pt}{1.2em}%
 Model &\ $\left(\frac{\sigma(pp\to h)^{\rm SM}}{\sigma(pp\to h)}\frac{\Gamma_{\rm tot}}{\Gamma_{\rm tot}^{\rm SM}}\right)R_{\tau^+\tau^-}$   &  $X_{\mu^+\mu^-}/(m_\mu^2/m_\tau^2)$\ &\ $X_{\mu\tau}$\  \cr \hline
SM & $1$ & $1$ & $0$ \cr
NFC & $(V_{h\ell}^*v/v_\ell)^2$  & $1$ & $0$ \cr
MSSM & $(\sin\alpha/\cos\beta)^2$ & $1$ & $0$ \cr
MFV & $1+2av^2/\Lambda^2$ & $1-4bm_\tau^2/\Lambda^2$ & $0$ \cr
FN & $1+{\cal O}(v^2/\Lambda^2)$ & $1+{\cal O}(v^2/\Lambda^2)$ & ${\cal O}(|U_{23}|^2v^4/\Lambda^4)$ \cr
GL & $9$ & $25/9$ & ${\cal O}(X_{\mu^+\mu^-})$ \cr
\hline\hline
\end{tabular}
\end{center}
\end{table}

\section{Bounds on Yukawa couplings}
\label{sec:exp}

We consider the following mass basis Lagrangian:
\beq
{\cal L}_Y=-m_i\bar \ell_L^i \ell_R^i - \frac{Y_{ij}}{\sqrt2}\bar \ell_L^i \ell_R^j h+{\rm h.c.}\,,\ \ \ \ i,j=e,\mu,\tau\,.
\eeq
Our convention is such that within the SM, $Y_{ij}=\delta_{ij}\sqrt{2} m_i/v$ and $v=246$~GeV. We henceforth denote the diagonal couplings $Y_{ii}$ as $Y_i$.

\subsection{Indirect constraints}
\label{sec:ind}
In this subsection we describe the constraints on the leptonic Yukawa couplings from various loop processes: charged lepton radiative decays and decays into three charged leptons, and the electric and magnetic moments of the electron and the muon. Ref.~\cite{Harnik:2012pb} obtains upper bounds on the off-diagonal $Y_{ij}$ assuming that the diagonal ones assume their SM values. We restore the dependence of these constraints on the diagonal couplings, and present the upper bounds on the various couplings in Table~\ref{tab:ub}. All experimental bounds are taken from Ref.~\cite{PDG}. Note that our couplings convention differs from that of~\cite{Harnik:2012pb}, $(Y_{ij})^{\rm HKZ}=Y_{ij}/\sqrt{2}$.

In the $\mu \to e\gamma$, $\tau \to e \gamma$ and $\tau\to \mu \gamma$ processes, we have included both the one- and the two-loop contributions. We define the following ratios, related to the respective contributions:
\beq
r_{\ell}\equiv \left(\frac{m_\tau}{m_\ell}\right)\frac{12\left(-0.082\, \frac{Y_{t}}{\sqrt2}+0.11\right)}{\left(-4+3\log\frac{m_h^2}{m_{\ell}^2}\right)}\frac{m_h^2}{(125\, {\rm GeV})^2}\,,\ \ \ \ \ell=\tau,\mu\,,
\eeq
and
\beq
r_{ij}\equiv \frac{Y_{i}}{Y_{j}^*+\sqrt 2 r_j}\,,\ \ \ \
\alpha_{ij}\equiv\frac{-4+3\log\frac{m_h^2}{m_i^2}}{-4+3\log\frac{m_h^2}{m_j^2}}\,,\ \ \ \ i,j=e,\mu,\tau\,.
\eeq
Using $Y_{t}=(Y_{t})_{\rm SM}=\sqrt{2}\bar m_t/v\approx0.95$ (with $\bar m_t=164$~GeV the top mass in the $\overline{\rm MS}$ renormalization scheme), one obtains $r_\mu\approx 0.27$ and $r_\tau\approx 0.03$. Numerically, $\alpha_{\mu\tau}\approx 1.76$, $\alpha_{e\tau}\approx3.3$, $\alpha_{e\mu}\approx 1.9$, and in the SM $r_{\mu\tau}^{\rm SM}\approx 0.01$, $r_{e\tau}^{\rm SM}\approx5\times 10^{-5}$, and $r_{e\mu}^{\rm SM}\approx1\times 10^{-4}$. The following bounds hold:
\beq
\tau\to \mu \gamma:\ \ \ |Y_{\tau}+\sqrt{2}r_\tau|\sqrt{(|Y_{\tau\mu}|^2+|Y_{\mu\tau}|^2)(1+|\alpha_{\mu\tau}|^2|r_{\mu\tau}|^2)+4\alpha_{\mu\tau}{\rm Re}[r_{\mu\tau}Y_{\tau\mu}^*Y_{\mu\tau}^*]}&<&1.2\times10^{-3}\,\no\\
\tau\to e \gamma:\ \ \ |Y_{\tau}+\sqrt{2}r_\tau|\sqrt{(|Y_{\tau e}|^2+|Y_{e\tau}|^2)(1+|\alpha_{e\tau}|^2|r_{e\tau}|^2)+4\alpha_{e\tau}{\rm Re}[r_{e\tau}Y_{\tau e}^*Y_{e\tau}^*]}&<&1.1\times10^{-3}\,\no\\
\mu\to e \gamma:\ \ \ |Y_{\mu}+\sqrt{2}r_\mu|\sqrt{(|Y_{\mu e}|^2+|Y_{e\mu}|^2)(1+|\alpha_{e\mu}|^2|r_{e\mu}|^2)+4\alpha_{e\mu}{\rm Re}[r_{e\mu}Y_{\mu e}^*Y_{e\mu}^*]}&<&1.9\times10^{-6}\,.\no\\
\eeq
In the SM, the $|\alpha_{ij}|^2 |r_{ij}|^2$ term is at most ${\cal O}(10^{-4})$ and completely negligible. The $\alpha_{ij}r_{ij}$ term is of order $\sim 0.1$ for $\tau\to \mu\gamma$ and less than a permil for $\tau\to e\gamma$ and $\mu\to e\gamma$, and thus subdominant. In contrast, the two-loop effect, manifested in the presence of $r_{\mu}$ and $r_{\tau}$, dominates over the one loop effect in $\tau\to \mu \gamma$ and $\mu\to e\gamma$, and in $\tau\to e \gamma$ both one- and two-loop effects give comparable contributions. In Table~\ref{tab:ub} we keep simplified expressions, obtained in the $r_{ij}\to 0$ limit. The processes $\mu\to 3e$, $\tau\to 3\mu$ and $\tau\to 3e{\rm \ or}\ e\mu \mu$ place bounds on the same combinations of Yukawa couplings as $\mu\to e\gamma$, $\tau\to e\gamma$ and $\tau\to \mu\gamma$ respectively. The bounds from the decays into three leptons are weaker by about an order of magnitude than those from the corresponding radiative decays.  In Table~\ref{tab:ub} we quote the strongest among these bounds.

\begin{table}[t]
\caption{Upper bounds on various Yukawa couplings}
\label{tab:ub}
\begin{center}
\begin{tabular}{ccc} \hline\hline
\rule{0pt}{1.2em}%
 Process & Yukawa couplings \ &  Upper bound \cr \hline
$\mu\to e\gamma$ & $|Y_{\mu}+\sqrt2\, r_\mu|\sqrt{|Y_{\mu e}|^2+|Y_{e\mu}|^2}$ & $1.9\times10^{-6}$ \cr
$\tau\to e\gamma$ & $|Y_{\tau}+\sqrt2\, r_\tau|\sqrt{|Y_{\tau e}|^2+|Y_{e\tau}|^2}$ & $1.1\times10^{-3}$ \cr
$\tau\to \mu\gamma$ & $|Y_{\tau}+\sqrt2\,r_\tau|\sqrt{|Y_{\tau \mu}|^2+|Y_{\mu\tau}|^2}$ & $1.2\times10^{-3}$ \cr
$(g-2)_e$ & Re($Y_{\mu e}Y_{e\mu})$ & $-0.040...0.055$ \cr
$(g-2)_e$ & Re($Y_{\tau e}Y_{e\tau})$ & $[-4.3...5.8]\times 10^{-3}$ \cr
$(g-2)_\mu$ &  Re($Y_{\tau\mu}Y_{\mu\tau}$)  & $(5.7\pm1.6)\times10^{-3}$ \cr
$(g-2)_\mu$ & Re($Y_{\mu}^2$)  & $(5.2\pm1.4)\times10^{-2}$ \cr
$d_e$ & $|{\rm Im}(Y_{\tau e}Y_{e \tau})|$ & $2.2 \times 10^{-8}$ \cr
$d_e$ & $|{\rm Im}(Y_{\mu e}Y_{e \mu})|$ & $2.0 \times 10^{-7}$ \cr
$d_e$ & $|{\rm Im}(Y_{e}^2)|$ & $2.2 \times 10^{-5}$ \cr
\hline\hline
\end{tabular}
\end{center}
\end{table}

As concerns $a_\mu=(g-2)_\mu$, the entry in the Table assumes that the discrepancy between the SM calculation and the experimental result~\cite{Bennett:2006fi} is accounted for by the respective $h$-related contribution:
\beqa\label{eq:amu}
a_\mu^h&=&\frac{{\rm Re}(Y_{\mu}^2)}{16\pi^2} \frac{m_\mu^2}{m_h^2}\left(\log \frac{m_h^2}{m_\mu^2}-\frac76\right)\nonumber\\
&+&\frac{{\rm Re}(Y_{\mu\tau}Y_{\tau\mu})}{16\pi^2} \frac{m_\mu m_\tau}{m_h^2}\left(\log \frac{m_h^2}{m_\tau^2}-\frac32\right)\no\\
&+&\frac{{\rm Re}(Y_{\mu e}Y_{e\mu})}{16\pi^2} \frac{m_\mu m_e}{m_h^2}\left(\log \frac{m_h^2}{m_e^2}-\frac32\right).
\eeq
Hence, in the corresponding entries we quote ranges, rather than an upper bound. If treated as upper bounds, and if diagonal Yukawa couplings are assumed to be not much smaller than their SM values, we have
\beqa\label{eq:ub}
\sqrt{{\rm Re}(Y_{\mu}^2)}&\lsim&0.2,\\
\sqrt{{\rm Re}(Y_{\mu\tau} Y_{\tau\mu})}|&\lsim&0.07.
\eeqa
An expression similar to Eq.~\eqref{eq:amu} holds for $(g-2)_e$ with the appropriate Yukawa coupling replacements. Note that we do not quote $(g-2)_\mu$ and $(g-2)_e$ as bounding Re$(Y_{\mu e}Y_{e\mu}$) and Re($Y_{e}^2$) respectively, since the resulting bounds are ${\cal O}(1)$.

The imaginary parts of the various Yukawa couplings are constrained by experimental bounds on the muon and electron electric dipole moments. For example, the contribution to the electron electric dipole moment arising from one-loop processes with an internal tau, muon and electron is:
\beqa\label{eq:de}
d_e&=&-\frac{{\rm Im}(Y_{e}^2)}{32\pi^2} \frac{e\, m_e}{m_h^2}\left(\log \frac{m_h^2}{m_e^2}-\frac76\right)\nonumber\\
&-&\frac{{\rm Im}(Y_{\tau e}Y_{e\tau})}{32\pi^2} \frac{e\,m_\tau}{m_h^2}\left(\log \frac{m_h^2}{m_\tau^2}-\frac32\right)\no\\
&-&\frac{{\rm Im}(Y_{\mu e}Y_{e\mu})}{32\pi^2} \frac{e\,m_\mu}{m_h^2}\left(\log \frac{m_h^2}{m_\mu^2}-\frac32\right).
\eeq
The resulting bounds are quoted in Table~\ref{tab:ub}. Note that we do not quote the corresponding bounds on ${\rm Im}\left(Y_{\tau \mu}Y_{\mu \tau}\right)$, ${\rm Im}\left(Y_{\mu}^2\right)$ and ${\rm Im}\left(Y_{\mu e}Y_{e \mu}\right)$ from the $d_\mu$ measurements since they are of order $\mathcal{O}(1)$ or weaker.

\subsection{Direct searches}
\label{sec:dir}
We now describe the current experimental status and future prospects at the LHC regarding measurements of the Yukawa couplings of the lepton sector.

At ATLAS and CMS, the search for a SM higgs decaying to $\tau\tau$ is divided into 3 channels according to the $\tau$ decays: fully hadronic (`hadhad'), semi-leptonic (`lephad') and fully leptonic (`leplep'). Although the $\tau$ decays mainly hadronically, with a branching fraction of $\sim65\%$, the hadronic channel suffers from $\tau$ reconstruction problems. The leptonic decay is suppressed compared to the hadronic one, and is also accompanied by more MET due to the presence of more neutrinos, making the mass reconstruction harder (and increasing difficulty with increased peak luminosity). The presence of all channels in the analysis is important due to their various strengths and weaknesses. Each channel, characterized by the decay mode, includes several categories according to the production mechanism. The largest cross section is for gluon-gluon fusion (ggF), but the clear topology of the vector-boson fusion (VBF) production mode, of two forward jets, favors this category. Combining all production categories, in Table~\ref{tab:atlascms} we present the current experimental status in ATLAS~\cite{ATLAStau} and CMS~\cite{CMStau} of the $\tau\tau$ search on the production cross section times branching ratio normalized to the SM.

In the dimuon channel, until recently, a search for $h\to \mu\mu$ was only done within the MSSM neutral Higgs searches at ATLAS~\cite{Aad:2012yfa} and CMS~\cite{CMSmu}. Although this channel can be fully reconstructed, the small branching fraction of ${\cal O}(10^{-4})$ disfavored this channel in the SM searches. (For early work on this search, see \cite{Plehn:2001qg,Cranmer:2006zs}.) The discovery of the Higgs-like boson has triggered searches for $h\to \mu\mu$, focusing on the inclusive search using background modeling along the lines of the $h\to \gamma\gamma$ search. In Table~\ref{tab:atlascms} we present the current bounds on the production cross section times branching ratio in the dimuon channel, normalized to the SM. In interpreting the results of~\cite{Aad:2012yfa,CMSmu} we use the SM production cross section and branching ratios for a $125$~GeV Higgs~\cite{Dittmaier:2011ti}. In the near future, the analysis can benefit from specific searches in the different categories such as ggF, VBF and associated production (VH).

\begin{table}[t]
\caption{Current upper bounds at $95\%$ CL on $R_{\tau^+\tau^-}$ and $R_{\mu^+\mu^-}$ at ATLAS and CMS.}
\label{tab:atlascms}
\begin{center}
\begin{tabular}{c|ccc} \hline\hline
\rule{0pt}{1.2em}%
 Channel &\ $\sqrt s$ & Collected data  &\ Current bound \cr
  &\ [TeV]   &  [fb$^{-1}$] \cr \hline
$\tau\tau$ hadhad   & 7+8 & 4.6+13  & 2.6~\cite{ATLAStau} \cr
$\tau\tau$ lephad   & 7+8 & 4.6+13  & 2.0~\cite{ATLAStau} \cr
$\tau\tau$ leplep   & 7+8 & 4.6+13  & 5.7~\cite{ATLAStau} \cr \hline
$\tau\tau$ combined & 7+8 & 4.6+13  & 1.9~\cite{ATLAStau} \cr
                    & 7+8 & 17  & 1.63~\cite{CMStau} \cr \hline
$\mu\mu$            & 7   & 4.7-4.8 & 30~\cite{Aad:2012yfa} \cr
                    & 7   & 4.96 & 34~\cite{CMSmu}   \cr
\hline\hline
\end{tabular}
\end{center}
\end{table}

These results from direct LHC searches place an upper bound on $Y_{\tau}$, and an upper bound on $Y_{\mu}$ stronger than Eq.~(\ref{eq:ub}):
\beqa\label{eq:dup}
|Y_{\tau}|&\lsim &1.3\, Y_{\tau}^{\rm SM}\approx 1.3\times 10^{-2}\,,\nonumber\\
|Y_{\mu}|&\lsim &5.5\, Y_{\mu}^{\rm SM}\approx 3.3\times 10^{-3}\,.
\eeqa

Next we estimate the future prospects for the $h\to \tau\tau$ and $h\to \mu\mu$ search channels. We use the results of~\cite{ATLASfuture} for these two channels to estimate how strong the expected 95$\%$ CL upper limit on $R_{\tau^+\tau^-}$ and $R_{\mu^+\mu^-}$ will be in the absence of a signal as well as to learn at what luminosity discovery in these channels will be possible.

Ref.~\cite{ATLASfuture} quotes for $\sqrt s=14$~TeV with ${\cal L}_0=300$~fb$^{-1}$
\beq
\left({\Delta R}/{R_{\rm SM}}\right)_{\mu^+\mu^-}(@{\cal L}_0)&=&0.52\,,\no\\
\left({\Delta R}/{R_{\rm SM}}\right)_{\tau^+\tau^-}(@{\cal L}_0)&=&0.15\,.
\eeq
(The result for the $\tau\tau$ channel includes the results at $300$~fb$^{-1}$ that are extrapolated from all the $7+8$~TeV $\tau\tau$ searches.) In the absence of a signal, these approximately translate to 95$\%$ CL upper limits on $R_X$:
\beq
R_{\mu^+\mu^-}&\lsim& 1\,,\no\\
R_{\tau^+\tau^-}&\lsim& 0.3\,.
\eeq
The square root of this gives the expected upper limit at $95\%$ CL at 14 TeV with 300~fb$^{-1}$ on the relevant Yukawa couplings:
\beq
|Y_{\mu}|&\lsim& Y_{\mu}^{\rm SM}\approx6\times 10^{-4}\,,\no\\
|Y_{\tau}|&\lsim& 0.54\,Y_{\tau}^{\rm SM}\approx5.5\times 10^{-3}\,.
\eeq
In addition, the precision of the measurement of the ratio $X_{\mu^+\mu^-}$ is expected to approach~$60\%$.

Alternatively, we can ask how much integrated luminosity ${\cal L}_1$ at 14 TeV is required so that a signal that is at least $3\sigma$ away from zero is observed. Assuming the error bars in each channel are dominantly due to statistics, we scale the results of~\cite{ATLASfuture} by $\sqrt{\frac{{\cal L}_0}{{\cal L}_1}}(\Delta R/R)_0=1/3$, where $(\Delta R/R)_0\equiv (\Delta R/R)(@{\cal L}_0)$. We then find that ${\cal L}_1\approx60$~fb$^{-1}$ in the $\tau\tau$ channel suffices for discovery, while in the $\mu\mu$ channel one needs~$\sim750$~fb$^{-1}$.

\section{Multi Higgs Doublet Models with NFC}
\label{sec:nfc}
Consider a multi Higgs doublet model (MHDM), where only one of the doublets couples to the charged lepton sector. Such a feature within the framework of MHDM's is known as natural flavor conservation (NFC) \cite{Glashow:1976nt,Paschos:1976ay}. For simplicity we consider here NFC with renormalizable couplings and at tree level only, and reserve a more complete study to future work. Let us denote the scalar doublet that couples to the charged leptons as $\phi_\ell$, and its VEV by $v_\ell$. The scalar mass eigenstate $h$ that has been observed at ATLAS/CMS is some combination of the neutral components of all scalar doublets:
\beq
h=\sum_i V_{hi} \phi_i^0,
\eeq
where $\sum_i|V_{hi}|^2=1$. Then, the couplings of $h$ to charged lepton pairs is given by
\beq
Y_{ij}= V_{h\ell}^*\delta_{ij}\sqrt2 m_i/v_\ell.
\eeq
We learn the following points about the Higgs-related lepton flavor parameters in this class of models:
\begin{enumerate}
\item $h$ has no flavor off-diagonal couplings:
\beq\label{eq:nfcfc}
Y_{\mu\tau},Y_{\tau\mu}=0.
\eeq
\item The values of the diagonal couplings deviate from their SM values, for example,
\beq\label{eq:nfcfd}
Y_\tau=\frac{V_{h\ell}^*v}{v_\ell}\ \frac{\sqrt2 m_\tau}{v}.
\eeq
The second factor on the RHS is the SM tau-Yukawa coupling, while the first factor is a lepton flavor universal coefficient. In the two Higgs doublet model (2HDM) of type II (for a review, see \cite{Branco:2011iw}) and, in particular, in  the minimal supersymmetric standard model (MSSM) at tree level, Eq.~(\ref{eq:nfcfd}) assumes the familiar form,
\beq
Y_\tau=-\left(\sin\alpha/\cos\beta\right)Y_{\tau}^{\rm SM}.
\eeq
Radiative corrections due to supersymmetric particles are known and are in general small.
\item The ratio between the Yukawa couplings to different charged lepton flavor is the same as in the SM:
\beq\label{eq:nfcr}
Y_\mu/Y_\tau=m_\mu/m_\tau.
\eeq
\end{enumerate}

We conclude that if experiments establish $X_{\mu^\pm\tau^\mp}\neq0$ and/or $X_{\mu^+\mu^-}\neq(m_\mu/m_\tau)^2$,
NFC will be excluded. On the other hand, if experiments find no evidence for flavor off-diagonal $h$ decays, and establish
that $X_{\mu^+\mu^-}$ is close to its SM value, while $R_{\tau^+\tau^-}\neq1$, the idea of NFC will be strongly supported.
These conclusions are presented in the NFC row of Table \ref{tab:rxx}. The MSSM is a specific model within this framework. Its predictions are presented in the MSSM row of this Table.

\section{A Single Higgs Doublet with MFV}
\label{sec:mfv}
With a single Higgs, $Y_{ij}\neq\frac{\sqrt{2} m_i}{v}\delta_{ij}$ is a consequence of higher-dimensional operators. The leading ones are the $d=6$ terms. In the interaction basis, we have
\beq\label{eq:dsix}
{\cal L}_Y^{d=4}&=&-\lambda_{ij}\overline{L_i} E_j \phi+{\rm h.c.},\\
{\cal L}_Y^{d=6}&=&-\frac{\lambda^\prime_{ij}}{\Lambda^2}\overline{L_i} E_j \phi
(\phi^\dagger \phi)+{\rm h.c.}\,,\nonumber
\eeq
where $L$ and $E$ stand for $SU(2)$ doublet- and singlet-leptons respectively, and expanding around the vacuum we have $\phi=(v+h)/\sqrt2$. Defining $V_{L,R}$ via
\beq\label{eq:vlr}
\sqrt2 m=V_L\left(\lambda+\frac{v^2}{2\Lambda^2}\lambda^\prime\right)
V_R^\dagger v,
\eeq
where $m={\rm diag}(m_e,m_\mu,m_\tau)$, and defining $\hat\lambda$ via
\beq\label{eq:deflamh}
\hat\lambda=V_L\lambda^\prime V_R^\dagger,
\eeq
we obtain
\beq\label{eq:yneqm}
Y_{ij}=\frac{\sqrt2 m_i}{v}\delta_{ij}+\frac{v^2}{\Lambda^2}\hat\lambda_{ij}.
\eeq

To proceed, one has to make assumptions about the structure of $\hat\lambda$. In what follows, we adopt the assumption of minimal flavor violation (MFV) \cite{D'Ambrosio:2002ex}.

MFV requires that the leptonic part of the Lagrangian is invariant under an $SU(3)_L\times SU(3)_E$ global symmetry, with the left-handed lepton doublets transforming as $(3,1)$,
the right-handed charged lepton singlets transforming as $(1,3)$ and the charged lepton Yukawa matrix $Y$ is a spurion transforming as $(3,\bar3)$.

Specifically, MFV means that, in Eq. (\ref{eq:dsix}),
\beq
\lambda^\prime=a\lambda+b\lambda\lambda^\dagger\lambda+{\cal O}(\lambda^5),
\eeq	
where $a$ and $b$ are numbers.  Note that, if $V_L$ and $V_R$ are the diagonalizing matrices for $\lambda$, $V_L \lambda V_R^\dagger=\lambda^{\rm diag}$, then they are also the diagonalizing matrices for $\lambda\lambda^\dagger\lambda$, $V_L\lambda\lambda^\dagger\lambda V_R^\dagger=(\lambda^{\rm diag})^3$. Then, Eqs. (\ref{eq:vlr}), (\ref{eq:deflamh}) and (\ref{eq:yneqm}) become
\beq\label{eq:mfv}
\frac{\sqrt2m}{v} &=&\left(1+\frac{av^2}{2\Lambda^2}\right)\lambda^{\rm diag}
 +\frac{bv^2}{2\Lambda^2}(\lambda^{\rm diag})^3,\nonumber\\
\hat\lambda&=&a\lambda^{\rm diag}+b(\lambda^{\rm diag})^3
=a\frac{\sqrt2 m}{v}+\frac{2\sqrt2bm^3}{v^3},
\nonumber\\
Y_{ij}&=&\frac{\sqrt2m_i}{v}\delta_{ij}\left[1+\frac{a v^2}{\Lambda^2}
+\frac{2b m_i^2}{\Lambda^2}\right],
\eeq
where, in the expressions for $\hat\lambda$ and $Y$, we included only the leading universal and leading non-universal corrections to the SM relations.

We learn the following points about the Higgs-related lepton flavor parameters in this class of models:
\begin{enumerate}
\item $h$ has no flavor off-diagonal couplings:
\beq\label{eq:nfcfc}
Y_{\mu\tau},Y_{\tau\mu}=0.
\eeq
\item The values of the diagonal couplings deviate from their SM values. The deviation is small, of order $v^2/\Lambda^2$:
\beq\label{eq:mfvfd}
Y_\tau\approx\left(1+\frac{av^2}{\Lambda^2}\right)\  \frac{\sqrt2 m_\tau}{v}.
\eeq
\item The ratio between the Yukawa couplings to different charged lepton flavors deviates from its SM value. The deviation is, however, very small, of order $m_\ell^2/\Lambda^2$:
\beq\label{eq:mfvr}
\frac{Y_\mu}{Y_\tau}=\frac{m_\mu}{m_\tau}\left(1-\frac{2b(m_\tau^2-m_\mu^2)}{\Lambda^2}\right).
\eeq
\end{enumerate}

The predictions of the SM with MFV non-renormalizable terms are given in the MFV row of Table \ref{tab:rxx}.

\section{A Single Higgs Doublet with FN}
\label{sec:fn}
An attractive explanation of the smallness and hierarchy in the Yukawa couplings is provided by the Froggatt-Nielsen (FN) mechanism~\cite{Froggatt:1978nt}. In this framework, a $U(1)_H$ symmetry, under which different generations carry different charges, is broken by a small parameter $\epsilon_H$. Without loss of generality, $\epsilon_H$ is taken to be a spurion of charge $-1$. Then, various entries in the Yukawa mass matrices are suppressed by different powers of $\epsilon_H$, leading to smallness and hierarchy.

Specifically for the leptonic Yukawa matrix, taking $h$ to be neutral under $U(1)_H$, $H(h)=0$, we have
\beq
 \lambda_{ij}\propto\epsilon_H^{H(E_j)-H(L_i)}\,.
\eeq
We emphasize that the FN mechanism dictates only the parametric suppression. Each entry has an arbitrary order one coefficient.
The resulting parametric suppression of the masses and leptonic mixing angles is given by~\cite{Grossman:1995hk}
\beq
m_{\ell_i}/v\sim\epsilon_H^{H(E_i)-H(L_i)}\,,\ \ \ |U_{ij}|\sim\epsilon_H^{H(L_j)-H(L_i)}\,.
\eeq

Since $H(\phi^\dagger\phi)=0$, the entries of the matrix $\lambda^\prime$ have the same parametric suppression as the corresponding entries in $\lambda$ \cite{Leurer:1993gy}, though the order one coefficients are different:
\beq
\lambda^\prime_{ij}={\cal O}(1)\times\lambda_{ij}.
\eeq
This structure allows us to estimate the entries of $\hat\lambda_{ij}$ in terms of physical observables:
\beqa
\hat\lambda_{33}&\sim&m_\tau/v,\nonumber\\
\hat\lambda_{22}&\sim&m_\mu/v,\nonumber\\
\hat\lambda_{23}&\sim&|U_{23}|(m_\tau/v),\nonumber\\
\hat\lambda_{32}&\sim&(m_\mu/v)/|U_{23}|.
\eeqa

We learn the following points about the Higgs-related lepton flavor parameters in this class of models:
\begin{enumerate}
\item $h$ has flavor off-diagonal couplings:
\beqa\label{eq:fnfc}
Y_{\mu\tau}&=&{\cal O}\left(\frac{|U_{23}|v m_\tau}{\Lambda^2}\right),\nonumber\\
Y_{\tau\mu}&=&{\cal O}\left(\frac{v m_\mu}{|U_{23}|\Lambda^2}\right).
\eeq
\item The values of the diagonal couplings deviate from their SM values:
\beq\label{eq:fnfd}
Y_\tau\approx\frac{\sqrt2 m_\tau}{v}\ \left[1+{\cal O}\left(\frac{v^2}{\Lambda^2}\right)\right].
\eeq
\item The ratio between the Yukawa couplings to different charged lepton flavors deviates from its SM value:
\beq\label{eq:mfvr}
\frac{Y_\mu}{Y_\tau}=\frac{m_\mu}{m_\tau}\left[1+{\cal O}\left(\frac{v^2}{\Lambda^2}\right)\right].
\eeq
\end{enumerate}

The predictions of the SM with non-renormalizable terms whose flavor structure is dictated by the FN mechanism are given in the FN row of Table \ref{tab:rxx}.

\section{Higgs-dependent Yukawa couplings}
\label{sec:gl}
It is possible, in principle, to relate the hierarchy in the charged fermion masses to a scenario where all renormalizable Yukawa couplings, except $Y_t$, vanish, but non-renormalizable terms generate the other Yukawa couplings suppressed by various powers of $(\phi^\dagger \phi/\Lambda^2)$~\cite{Giudice:2008uua}. In particular, for the two heavier generations of charged leptons, we have
\beqa
{\cal L}\supset-\frac{\lambda^\prime_{33}}{\Lambda^2}\overline{L_{3}}E_{3} \phi(\phi^\dagger \phi)-\frac{\lambda^{\prime}_{23}}{\Lambda^2}\overline{L_{2}}E_{3} \phi(\phi^\dagger \phi)
-\frac{\lambda^{\prime\prime}_{22}}{\Lambda^4}\overline{L_{2}}E_{2} \phi(\phi^\dagger \phi)^2-\frac{\lambda^{\prime\prime}_{32}}{\Lambda^4}\overline{L_{3}}E_{2} \phi(\phi^\dagger \phi)^2+{\rm h.c.}
\eeqa
The powers of the diagonal terms are proposed in Ref.   \cite{Giudice:2008uua} in correspondence to the charged lepton masses (with $v^2/\Lambda^2\simeq1/60$), while the off-diagonal terms are our own ansatz, inspired by the value of $|U_{23}|$.

We learn the following points about the Higgs-related lepton flavor parameters in this class of models:
\begin{enumerate}
\item $h$ has flavor off-diagonal couplings:
\beqa\label{eq:glfc}
Y_{\tau\mu}&=&{\cal O}(Y_\mu)\,,\no\\
\frac{Y_{\mu\tau}}{Y_{\tau\mu}}&=&\frac{m_\mu}{m_\tau}\,.
\eeqa
\item The values of the diagonal couplings are larger by a large factor than their SM values:
\beq\label{eq:glfd}
Y_\tau\simeq3\ \frac{\sqrt2 m_\tau}{v}\,,\ \ \ Y_\mu\simeq5\ \frac{\sqrt2 m_\mu}{v}
\eeq
\item The ratio between the Yukawa couplings to different charged lepton flavors is larger than its SM value:
\beq\label{eq:glr}
\frac{Y_\mu}{Y_\tau}\simeq\frac53\ \frac{m_\mu}{m_\tau}.
\eeq
\end{enumerate}

The predictions of the SM with these Higgs-dependent Yukawa couplings are given in the GL row of Table~\ref{tab:rxx}. $R_{\tau^+\tau^-}$ is only mildly enhanced ($R_{\tau^+\tau^-}\sim 1.4$) relative to the strong enhancement of $|Y_{\tau}|^2$ due to the corresponding enhancements of $|Y_b|^2$ and $|Y_c|^2$. The model is excluded because it predicts a strong suppression of $R_{ZZ^*}$ and $R_{\gamma\gamma}$ which is inconsistent with collider measurements. This demonstrates the power of these measurements to exclude flavor models.

\section{Discussion and Conclusions}
\label{sec:con}
The discovery of a Higgs-like boson at the ATLAS and CMS experiments signals a new era, where the interplay between flavor physics and the physics of electroweak symmetry breaking will be explored for the first time. In this work, we demonstrate the power of various relevant measurements to improve our understanding of both the flavor sector and the electroweak breaking sector.

From the experimental side, we focus on measurements of various dilepton Higgs decay modes: $\tau^+\tau^-$, $\mu^+\mu^-$, $\tau^\pm\mu^\mp$. From the theoretical side, we select a few attractive models of the Higgs sector: the Standard Model, two Higgs doublet models with natural flavor conservation, and the Standard Model supplemented with non-renormalizable terms whose structure is either minimally flavor violating or subject to selection rules from a Froggatt-Nielsen mechanism. (In addition, we describe a model of Higgs-dependent Yukawa couplings that is already excluded by the Higgs data.)

Our results are summarized in Table \ref{tab:rxx} which shows that, in principle, these four frameworks can be distinguished by the measurements of the dilepton decays. Deviation of ${\rm BR}(h\to\tau\tau)$ from the SM ($R_{\tau^+\tau^-}\neq1$), accompanied with consistency of ${\rm BR}(h\to\mu^+\mu^-)/{\rm BR}(h\to\tau^+\tau^-)$ with the SM ($X_{\mu^+\mu^-}=m_\mu^2/m_\tau^2$) and the absence of flavor changing decays ($X_{\tau\mu}=0$) will be suggestive of natural flavor conservation. Violation of $R_{\tau^+\tau^-}=1$, a small violation of $X_{\mu^+\mu^-}=m_\mu^2/m_\tau^2$, and $X_{\tau\mu}=0$, will be suggestive of minimal lepton flavor violation. Comparable deviations from $R_{\tau^+\tau^-}=1$ and $X_{\mu^+\mu^-}=m_\mu^2/m_\tau^2$ and observation of small but finite rate for $h\to\tau\mu$ would suggest a less restrictive but still structured framework, such as the FN mechanism.

The various models that we analyzed show that, qualitatively, different flavor models make different predictions for the various flavor-related Higgs features. A more quantitative analysis, where we compare the experimental accuracy that can be hoped for with the size of effects expected in the various theoretical models, is in progress. We will also extend our work to additional new physics models, with potentially sizable effects, and to calculation of new physics loop effects on the Higgs decay rates into dileptons.

We are entering a new era of measuring the couplings of the newly discovered boson $h$. The interplay between flavor physics and the physics of electroweak symmetry breaking will provide an opportunity to achieve better understanding of both.

\vspace{1cm}
\begin{center}

{\bf Acknowledgements}
\end{center}
We thank Liron Barak and Shikma Bressler for many useful discussions, and Michael Spira and Eric Kuflik for helpful conversations and comments on the manuscript. YN is the Amos de-Shalit chair of theoretical physics. This project
is supported by the Israel Science Foundation and by the
German-Israeli foundation for scientific research and development (GIF).


\end{document}